\newcommand{\gm}{\ensuremath{\gamma}}
\newcommand{\fermi}{\textit{Fermi}}

\documentclass[a4paper,11pt]{article}
\usepackage{pos}
\usepackage{lineno}

\title{Statistical properties of flux variations in blazar light curves at GeV and TeV energies}
 \ShortTitle{Flux Variations in GeV and TeV Blazar Light Curves}
\manuallySeparateAuthors
\author*[a,b]{Sarah.~M.~Wagner,}
\author[a]{ P.~R.~Burd,}
\author[a,\#]{ D.~Dorner,}
\author[a,\#]{ K.~Mannheim;}

\author[a]{ S.~Buson,}
\author[a,c]{ A.~Gokus,}
\author[b]{ G. Madejski,}
\author[d]{ J.~D.~Scargle}
\author{ on behalf of the \textit{Fermi}-LAT Collaboration;}
\author[e]{ A. Arbet-Engels,}
\author[f]{ D.~Baack,}
\author[g]{ M.~Balbo,}
\author[e]{ A.~Biland,}
\author[e,h]{ T.~Bretz,} 
\author[f]{ J.~Buss,}
\author[f]{ D.~Elsaesser,}
\author[a]{ L.~Eisenberger,}
\author[e]{ D.~Hildebrand,}
\author[a]{ R.~Iotov,}
\author[a]{ A.~Kalenski,}
\author[e]{ D.~Neise,}
\author[f]{ M.~Noethe,}
\author[a]{ A.~Paravac,}
\author[f]{ W.~Rhode,}
\author[a]{ B.~Schleicher,}
\author[g]{ V.~Sliusar,}
\author[g]{ R.~Walter}
\author{ on behalf of the FACT Collaboration}

\affiliation[a]{Julius-Maximilians Universität Würzburg, Fakultät für Physik und Astronomie, Emil-Fischer-Straße 31, 97074 Würzburg, Germany}
\affiliation[b]{Kavli Institute for Particle Astrophysics and Cosmology and SLAC National Accelerator Laboratory, Stanford University, Menlo Park, California 94025, USA}
\affiliation[c]{Dr. Karl Remeis-Observatory and Erlangen Centre for Astroparticle Physics, Universit\"at Erlangen-N\"urnberg, Sternwartstr.~7, 96049 Bamberg, Germany}
\affiliation[d]{Astrobiology and Space Science Division, NASA Ames Research Center, Moffett Field, California 94035-1000, USA}
\affiliation[e]{ETH Zurich, Institute for Particle Physics and Astrophysics, Otto-Stern-Weg 5, 8093 Zurich, Switzerland}
\affiliation[f]{TU Dortmund, Experimental Physics 5, Otto-Hahn-Str. 4a, 44227 Dortmund, Germany}
\affiliation[g]{University of Geneva, Department of Astronomy, Chemin d\'Ecogia 16, 1290 Versoix, Switzerland}
\affiliation[h]{also at RWTH Aachen University}
\affiliation[\#]{also in FACT}

\emailAdd{sarah.wagner@physik.uni-wuerzburg.de}

\abstract{
Despite numerous detections of individual flares, the physical origin of the rapid variability observed from blazars remains uncertain.
Using Bayesian blocks and the Eisenstein-Hut HOP algorithm, we characterize flux variations of high significance in the \gm-ray light curves of two samples of blazars. Daily binned long-term light curves of TeV-bright blazars observed with the First G-APD Cherenkov Telescope (FACT) are compared to those of GeV-bright blazars observed with the Large Area Telescope on board the \fermi~Gamma-ray Space Telescope (\fermi-LAT). We find no evidence for systematic asymmetry of the flux variations based on the derived rise and decay time scales. Additionally, we show that the daily-binned blazar light curves can be described by an exponential stochastic Ornstein-Uhlenbeck process with parameters depending on energy. Our analysis suggests that the flux variability in both samples is a superposition of faster fluctuations. This is, for instance, challenging to explain by shock-acceleration but expected for magnetic reconnection.}

\FullConference{37$^{\rm{th}}$ International Cosmic Ray Conference (ICRC 2021)\\
		July 12th -- 23rd, 2021\\
		Online -- Berlin, Germany}

\begin{document}
\maketitle

\section{Introduction}
Blazars are active galactic nuclei (AGN) with highly relativistic outflows that are directed closely towards our line of sight. They can be sub-classified
based on their spectral energy distribution (SED), which characteristically shows a double-hump structure. The low-energy hump is well described by synchrotron emission while the high-energy hump is likely due to inverse Compton scattering and/or hadronic processes. Based on the position of the peak of the former, blazars can be divided in low- and high-synchrotron-peaked objects (LSP and HSP) \cite{Abdo_2010_SEDs}. 
There have been many attempts to link these observational properties to the physical nature of blazars \cite{Ghisellini_2017}. However, it remains uncertain which intrinsic differences there are between the blazar classes and what physical mechanisms accelerate the particles necessary to produce the observed $\gamma$-radiation in the first place \cite{Madejski_Sikora_2016}. We study the flux variability in the range of the high-energy hump in the brightest LSP and HSP objects in order to compare the blazar classes and constrain possible emission models.


\section{Data and instruments}
The high-energy hump of LSP and HSP objects falls in different energy ranges.
Currently, no instrument covers the wide \gm-ray range with continuous sensitivity. Hence, we utilize the two instruments described in the following to measure flux variations of blazars in daily binning at both lower as well as higher \gm-ray energies.

\paragraph{\fermi-LAT}
The Large Area Telescope (LAT) on board the \fermi~satellite operates from 20\,MeV up to 1\,TeV. Due to its capability of monitoring the entire sky within a day, it is possible to gain continuous information about each \gm-ray source over the whole length of the \fermi~mission.
Hence, we computed light curves with daily binning from 2008-08-05 to 2020-12-31 (MET 239587201 - 631065605) for an energy range between 100\,MeV and 300\,GeV. Our source sample for \fermi-LAT analysis consists of the brightest blazars in the Fourth \fermi-LAT source catalog (4FGL)\cite{fermi_4fgl}, as listed in Table\,\ref{tab:sources} on the left.

We performed the standard data reduction process\footnote{~\href{https://fermi.gsfc.nasa.gov/ssc/data/analysis/documentation/}{https://fermi.gsfc.nasa.gov/ssc/data/analysis/documentation/}} with \texttt{Science Tools 1.2.23},\texttt{fermipy 0.20.0}.
We model all sources included in the 4FGL catalog that are within a region of interest (ROI) of $15^\circ$ around each source, and model the isotropic diffusion emission with \texttt{iso\_P8R3\_SOURCE\_V2\_v1} and the Galactic diffuse emission with \texttt{gll\_iem\_v07}.
For further analysis of the light curves, we only consider flux bins with test statistic (TS)$>4$ and a flux uncertainty equal to or smaller than its flux, analogous to \cite{Meyer_2019}. 

\paragraph{FACT}
Since October 2011, the First G-APD Cherenkov Telescope (FACT) has been measuring $\gamma$-rays from hundreds of GeV up to several TeV. The three brightest blazars with continuous light curves observed by FACT are listed in Table\,\ref{tab:sources}. The data were taken from 2011-11-15 to 2020-01-23 and are analyzed using the Modular Analysis and Reconstruction Software (MARS) \cite{2010apsp.conf..681B}. For the background suppression, the 'light curve cuts' as described in \cite{2019ICRC...36..630B} were used. A data quality selection based on the artificial trigger rate \cite{2019ICRC...36..630B,2017ICRC...35..779H} was applied. More details on the analysis can be found in \cite{2021A&A...647A..88A}. 
Since the observations are ground-based, there are seasonal gaps of $\sim120$~days each year. We divide each FACT light curve into seven stretches where the source could be observed continuously\footnote{~Further short gaps on the order of days can be caused by bad weather and the break in observing around full moon.} and refer to these as (FACT light curve) chunks. 

\setlength{\tabcolsep}{10pt}
\begin{table}
\centering
\begin{tabular}{cc}
\begin{tabular}[t]{l|c|c}
\multicolumn{3}{c}{\fermi-LAT (all FSRQ and LSP)} \\
\hline
Name & $\mathrm{log}(\nu_{\text{syn}}/\mathrm{Hz})$ & $z$ \\
\hline 
\hline
3C\,454.3 & 13.6 & 0.859 \\ 
\hline 
PKS\,1510-089 & 13.1 & 0.36 \\ 
\hline 
3C\,279 & 12.6 & 0.536 \\ 
\hline 
PKS\,1424-41 & 13.1 & 1.522 \\ 
\hline 
PKS\,1830-211 & 12.2 & 2.5 \\ 
\hline 
4C\,+21.35 & 13.5 & 0.433 \\ 
\hline 
3C\,273 & 13.5 & 0.158 \\ 
\end{tabular} &
\begin{tabular}[t]{l|c|c}
\multicolumn{3}{c}{FACT (all BLL and HSP)} \\
\hline
Name & $\mathrm{log}(\nu_{\text{syn}}/\mathrm{Hz})$ & $z$ \\
\hline 
\hline
Mrk\,421 & 16.6 & 0.03 \\ 
\hline
Mrk\,501 & 17.1 & 0.033 \\
\hline
1ES\,1959+650 & 16.6 & 0.047 \\ 
\end{tabular} 
\tabularnewline
\end{tabular}
\caption{Sources considered for \fermi-LAT (left) and FACT (right) with the corresponding logarithm of the position of the synchrotron peak frequency $\nu_{\text{syn}}$ and the redshift $z$ taken from 4LAC \cite{4LAC_2019} and \cite{Abdo_2010_SEDs}.}
\label{tab:sources}
\end{table}

\section{Definition of flares and their asymmetry}
We derive the best fit piece wise constant representation for all \fermi-LAT light curves and FACT chunks by applying the Bayesian block algorithm\footnote{~\href{https://docs.astropy.org/en/stable/api/astropy.stats.bayesian_blocks.html}{https://docs.astropy.org/en/stable/api/astropy.stats.bayesian\_blocks.html}}\footnote{~Assuming that the flux errors are Gaussian, we set fitness to 'measures'.} \cite{Scargle_2013}. For the purpose of this work, we define flares as groups of blocks, following \cite{Meyer_2019}. The algorithms implemented for this procedure are available on GitHub\footnote{~Code for flare analysis: \href{https://github.com/swagner-astro/lightcurves}{https://github.com/swagner-astro/lightcurves}}.

\begin{figure}
\begin{center}
\includegraphics[width=\linewidth]{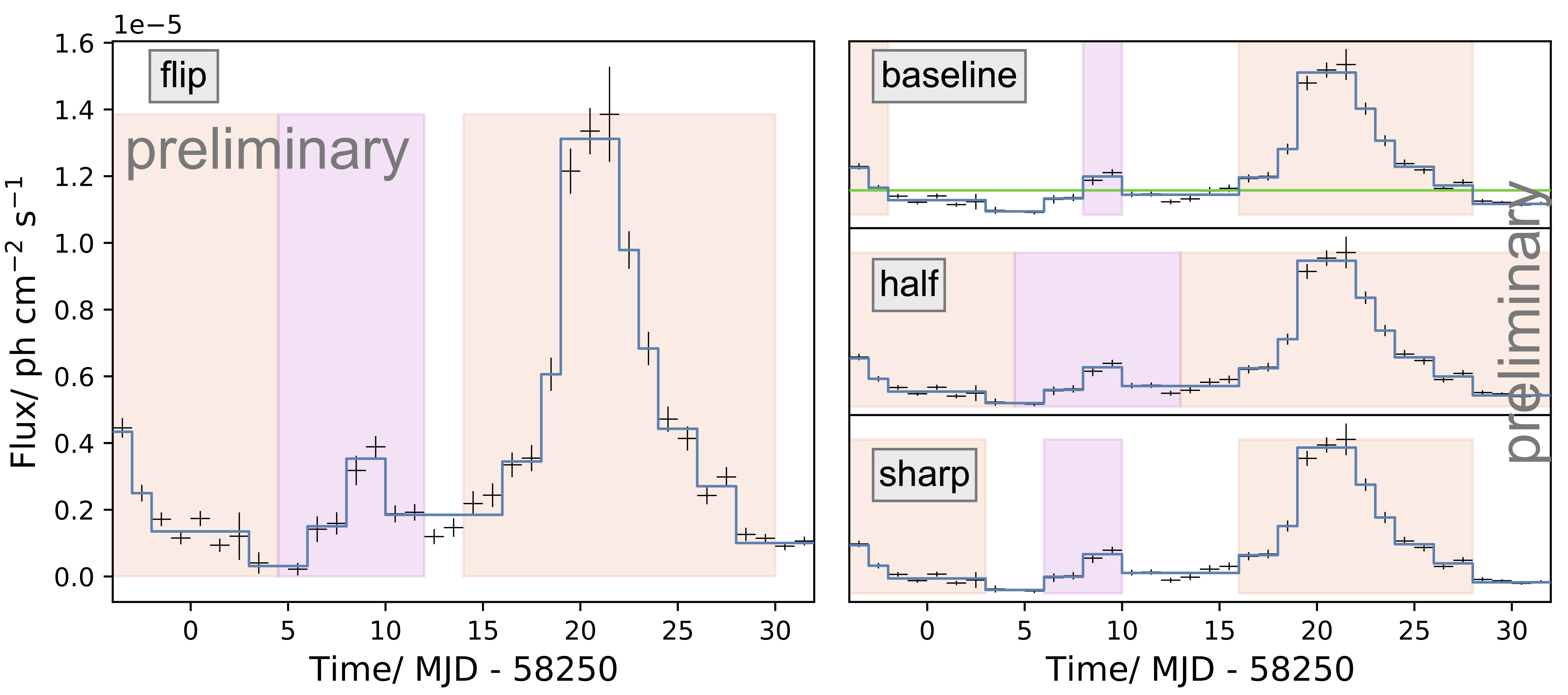}
\caption{All panels show an identical part of the daily-binned \fermi-LAT light curve of 3C\,279 (black bins) and its Bayesian block representation (blue line).
Flares defined with the corresponding method (top left of each panel) are shaded orange and purple. The green baseline in the top right panel shows the quiescent background introduced in \cite{Meyer_2019}.}
\label{fig:HOP}
\end{center}
\end{figure}

We determine the peak times to be at the center of local maxima in the block representation and every block that is subsequently lower to the left and right belongs to that peak (``hop\footnote{~Thus, referred to as HOP algorithm, which is not an acronym.}'' downwards). For the resulting ``valley'' blocks, we implemented four methods to define where one flare ends and another one starts. We illustrate those with an exemplary part of the \fermi-LAT light curve of 3C\,279 in Fig.\,\ref{fig:HOP}.

The \textit{half} and \textit{sharp} methods simply divide the valley block in half, or neglect it completely, respectively. 
The \textit{flip} method extrapolates the slope of the flare by flipping the length of the adjacent block onto the valley block. The start/end then sits at the bordering edge of the valley block minus/plus the length of the adjacent block, as shown in Fig.\,\ref{fig:HOP} on the left. 
In order to avoid overlap, we set the start/end of a flare to the center of the block (\textit{half}) if flipping the block would exceed the center of the valley block.
Originally, the start/end of the so-called HOP groups is determined by the flux exceeding/dropping under a predefined flux level (baseline, e.g. the quiescent level \cite{Meyer_2019}, see top panel of the right plot in Fig.\,\ref{fig:HOP}). We refer to this as \textit{baseline} method. In case of multiple flares above the baseline, we again divide the valley block in half (\textit{half}). While all of the methods have advantages and disadvantages, it becomes clear that an objective definition of a flare or flux variation in this context always needs some kind of assumption. For this work, we choose the \textit{flip} method since it allows to factor in as much data as possible without stretching the flares to be unnaturally long. The results do not change significantly, for either of the methods illustrated here. Based on the determined start, peak and end time of each flare, we compute the asymmetry measure
\begin{equation}
A = \frac{t_{\text{rise}} - t_{\text{decay}}}{t_{\text{rise}} + t_{\text{decay}}} 
\label{eq:asym}
\end{equation}
which can shed light on the physical mechanisms that cause the high-energy variability of blazars. A negative asymmetry indicates $t_{\text{rise}}<t_{\text{decay}}$ and can for instance be explained by a moving shock colliding with a stationary feature \cite{Sokolov_2004}. Symmetric flares can be explained by mini-jets sweeping across the line of sight with a constant angular velocity \cite{Giannios_2009}. Physical emission models producing flares with positive asymmetry ($t_{\text{rise}}>t_{\text{decay}}$) are less common.

\begin{figure}
\includegraphics[width=\linewidth]{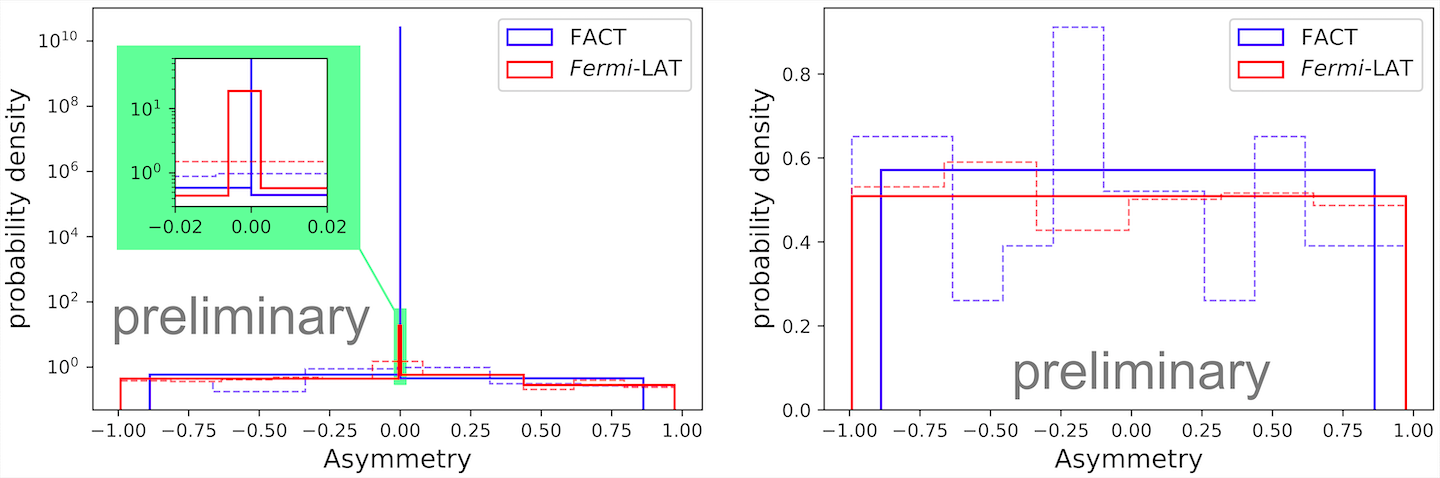}
\caption{Distribution of asymmetry measures for the \fermi-LAT (red) and FACT (blue) flares in Bayesian (solid thick line) and constant binning (thin dashed line); \textit{Left}: all flares; \textit{Right}: flares with at least two blocks.}
\label{fig:asym}
\end{figure}

A large portion of flares in the \fermi-LAT light curves as well as the FACT chunks are found to consist of just a single Bayesian block (47\% and 48\%, respectively). Such flares often result in a trivial asymmetry of exactly zero (\textit{flip} to both sides). Utilizing Bayesian blocks for binning\footnote{~Due to the variable bin width we divide the number of counts in each bin by its width, resulting in the probability density. This holds for all histograms in this work.} the distribution of asymmetry measures, as shown in Fig.\,\ref{fig:asym} on the left with thick solid lines, illustrates how prominent this feature is. Interestingly, this artefact is not at all captured with constant binning (thin dashed lines). Since single block flares do not allow the resolution of the true asymmetry of the flux variation, we exclude them for both samples. The asymmetry measures of all remaining flares are shown in the same figure on the right for \fermi-LAT in red and FACT in blue. In this case, the Bayesian binning (thick solid lines) results in one block confirming the visual impression of the constant binning (thin dashed lines) that every kind of asymmetry occurs to a comparable degree. Based on a two-sided Kolmogorov-Smirnov (KS) test\footnote{~\href{https://docs.scipy.org/doc/scipy/reference/generated/scipy.stats.ks_2samp.html}{https://docs.scipy.org/doc/scipy/reference/generated/scipy.stats.ks\_2samp.html}}, it can not be rejected that the asymmetry measures from \fermi-LAT and FACT are drawn from the same distribution. This holds for all flares (KS: p $\sim 0.42$) as well as the sub-sample with at least two blocks per flare (KS: p $\sim 0.97$). The distribution of asymmetry measures is not correlated with the (redshift corrected) duration of the flares nor their flux amplitude\footnote{~The Pearson and Spearman correlation coefficients show a range of $-0.17 < r_{\mathrm{Pearson/Spearman}} < 0.02 $}. Based on the method utilized here, we can not distinguish the asymmetry of the TeV (\fermi-LAT) and GeV (FACT) flux variations with daily binning. 


\section{Quantization of the variability using Ornstein-Uhlenbeck (OU) parameters}
In an additional approach, we interpret the flux variations as correlated noise parametrized by a first order auto-regressive process. We use the discrete OU process (see Eq.\,5e in \cite{Burd_2021})
\begin{equation}
    \label{Eq.Discrete_SDE}
    u_{T+1} = u_T + \theta\Delta t (\mu - u_T) + \sigma \sqrt{\Delta t} \mathcal{N}_T
\end{equation}
with mean revision level $\mu$, mean revision rate $\theta$, and white noise constituted by independent draws from a normal distribution $\mathcal{N}(m = 0, \sigma ^2)$. We interpret the value of the time series $u_T$ to be the logarithm of flux and extract the OU parameters $(\mu,\sigma,\theta)$ of each light curve\footnote{~The fact that we are able to identify HOP groups indicates stationarity of the analyzed light curves on relevant time-scales which is necessary for this procedure.} following \cite{Burd_2021}\footnote{~Code for OU paramter extraction: \href{https://github.com/PRBurd/astro-wue}{https://github.com/PRBurd/astro-wue}}.

\begin{figure}
\includegraphics[width=0.49\linewidth]{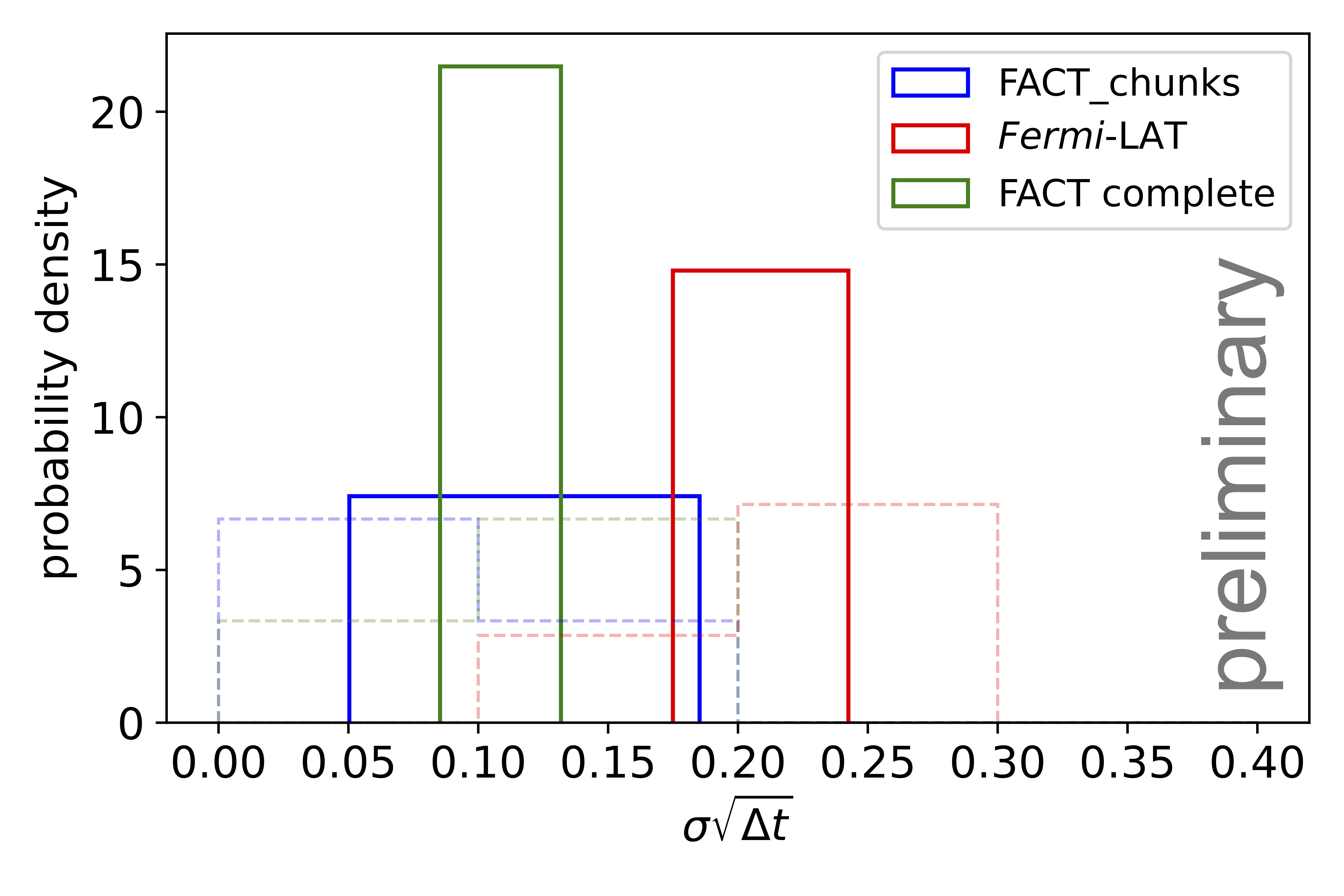}
\includegraphics[width=0.49\linewidth]{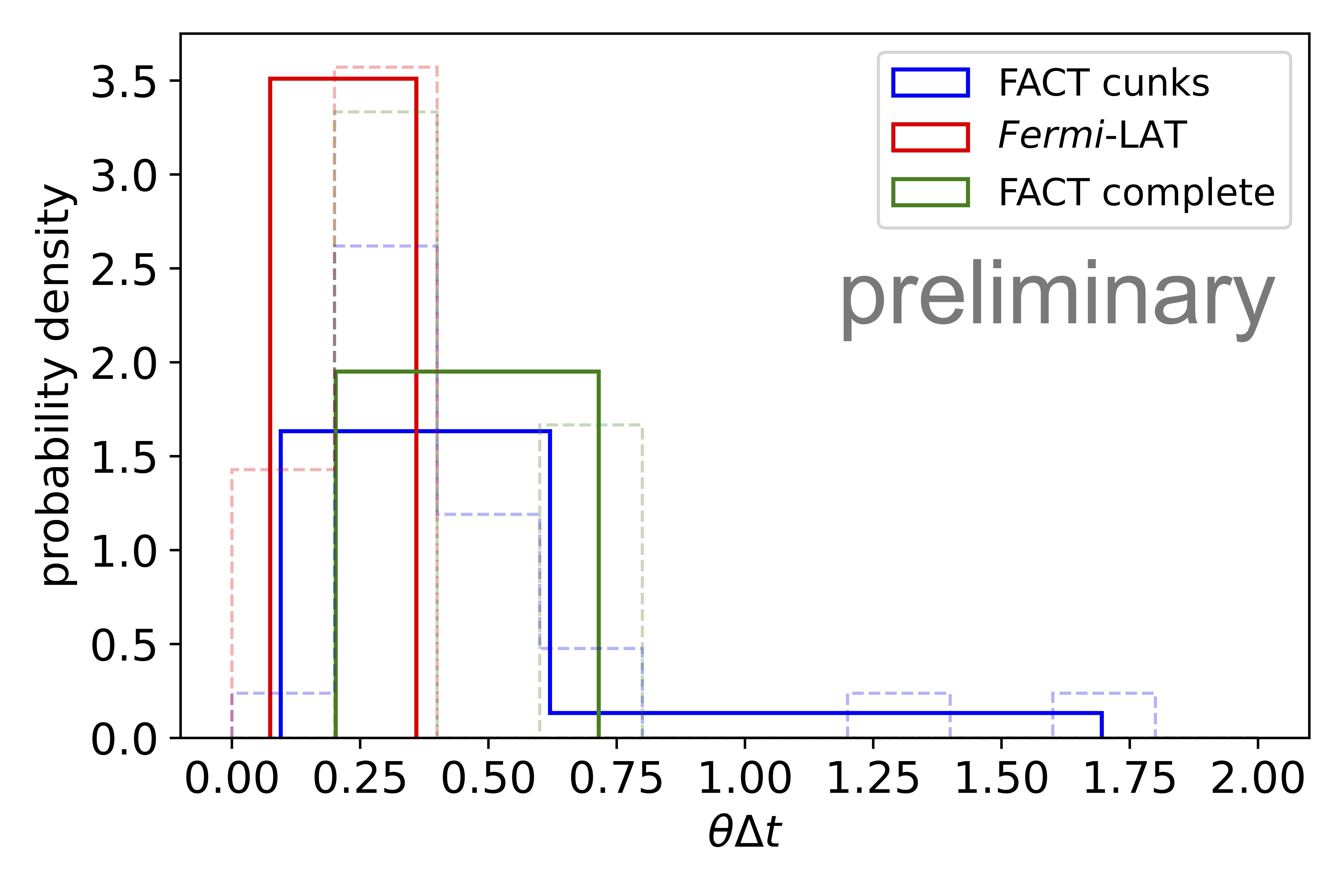}
\includegraphics[width=0.49\linewidth]{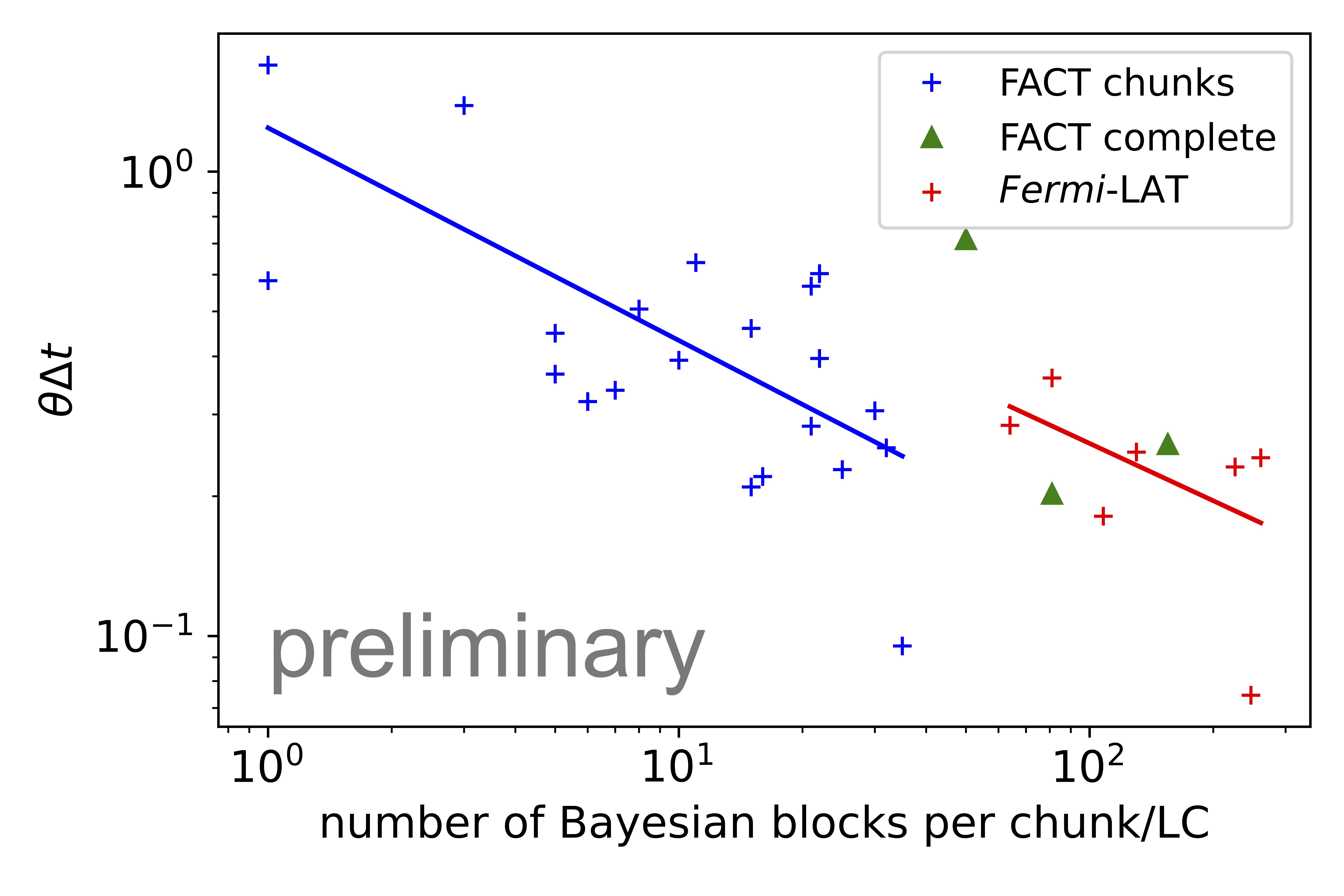}

\caption{\textit{Top-left:} Distribution of the $\sigma$-parameter; \textit{Top-right:} distribution of the $\theta$-parameter; \textit{Bottom-right:}The number of Bayesian blocks for a light curve against the $\theta$-parameters fit with a power law.}
\label{fig:OU_results0}
\end{figure}

Figure\, \ref{fig:OU_results0} (top-left) shows the resulting distribution of $\sigma$-parameters for the \fermi-LAT light curves (red), the complete FACT light curves (green) and the FACT chunks (blue) in Bayesian binning. The $\sigma$-value can be interpreted as the strength of the innovation, which corresponds to the maximum random amplitude. There are significant differences for this parameter among the three classes. A two-sided KS test yields a high probability that the \fermi-LAT sample is not drawn from the same distribution as the FACT samples (KS complete:$p\sim 0.017$, KS chunks: $p\sim 1.4\times 10^{-5}$). This indicates that the amplitude of random variability is smaller in the \fermi-LAT sample as compared to the FACT chunks and complete light curves.
Figure~\ref{fig:OU_results0} (top-right) shows the $\theta$-distribution for the \fermi-LAT light curves (red), the complete FACT light curves (green) and the FACT chunks (blue). This parameter denotes the reversion rate drawing back the flux variation to a mean reversion level. Smaller $\theta$-values indicate light curves (or segments in light curves) where variability takes place on time scales which are large with respect to the time binning, while larger $theta$-values indicate flux variations which quickly (with respect to the time-sampling) level back out to the reversion level. Also in this case, though on a smaller significance level, the \fermi-LAT light curves and the FACT light curve chunks are not drawn from the same distribution (KS chunks: $p\sim 0.05$). This does, however, not hold for the complete FACT light curves (KS complete:$p\sim 0.85$).

In Fig.~\ref{fig:OU_results0} (bottom-left) the number of Bayesian blocks in each light curve (chunk) is plotted against its $\theta$-parameter resulting in an anti-correlation\footnote{~
$R_{\mathrm{Pearson}}= -0.54/-0.58$ and 
$R_{\mathrm{Spearman}} = -0.57/-0.61$ for FACT chunks/\fermi-LAT respectively.}. As expected, the overall longer \fermi-LAT and complete FACT light curves consist of more blocks than the short FACT chunks. A power law of the form $\text{Number of block}\propto \theta^s$ is fit with the resulting slopes $s_{\mathrm{FACT chunks}} = -0.458 \pm 0.093$ and $s_{\mathrm{Fermi-LAT}} = -0.41 \pm 0.24$. Three data points are too few to draw conclusions for the complete FACT light curves but they do not contradict the measured trend. 

\section{Discussion}
Analyzing GeV (\fermi-LAT) and TeV (FACT) blazar light curves with Bayesian blocks and the HOP algorithm shows that a large fraction of the flux variations is best described with just one block. In this case the rise and decay are not resolved enough, suggesting that the true variability (possibly tracing back to the physical process) takes place on even smaller timescales than days. \fermi-LAT has, for instance, observed variability within several minutes \cite{Shukla2020} and FACT detected intranight variability for Mrk\,421 \cite{ATel_FACT2017}. Thus, it is very likely that the variations in the daily binned GeV and TeV light curves are superpositions of individual, smaller flux fluctuations. This is further supported by the similarity of our results to correspondingly analyzed \fermi-LAT light curves in monthly binning \cite{Burd_2021}. Both appear to be too large to resolve the high-energy blazar variability.
When excluding the single-block flares, every kind of asymmetry ($-1\leq A \leq1$) is present with the same probability for both samples, see Fig.\,\ref{fig:asym}. It is unlikely that a mechanism producing strongly positively asymmetric flares results in a light curve with just as many negatively as positively asymmetric flux fluctuations (on the same time scale). Possible theories for particle acceleration and emission models need to reproduce this variety of high-energy flux fluctuations in daily binning. $\gamma$-ray flares powered in one or several plasmoids provide a possible explanation \cite{Meyer_2021}.
The difference in the $\sigma$-distributions for the chunks and complete FACT light curves can be understood in the following way. If there are chunks which show large amplitude variations, the extracted $\sigma$-parameter will have a larger value. When extracting the $\sigma$-parameter of an entire FACT light curve, the extracted $\sigma$-value will be influenced by long-time behavior possibly containing very little variation in flux. The anti-correlation of the numbers of Bayesian blocks and the $\theta$-parameters, can be interpreted as follows. The smaller the $\theta$-parameter, the larger the number of Bayesian blocks describing the light curve (chunks). This means that the variability for these sources is more complex and over all larger. Our findings show that the $\theta$ and $\sigma$-parameters can be used to quantify the variability in a light curve.

We conclude that daily binning at the given sensitivities is insufficient to infer strong constraints on the physical emission mechanisms driving variability in LSP as well as HSP blazars. Emission models for \gm-ray radiation need to be able to reproduce the properties of the observed light curves, particularly the fast variability, analyzed here.


\section*{Acknowledgements}
SMW is grateful for the Klaus Murmann Scholarship awarded by the Stiftung der Deutschen Wirtschaft (sdw, German Business Foundation) as well as the warm hospitality and great inspiration at KIPAC/SLAC and Stanford University. PRB acknowledges {\em DATEV Stiftung Zukunft} for funding  the interdisciplinary data lab {\em DataSphere@JMUW}. JDS thanks Joe Bredekamp, the NASA Applied Information Systems Research Program and the NASA Astrophysics Data Analysis Program (Grant NNX16AL02G) for support.
\textbf{Fermi}: The \textit{Fermi}-LAT Collaboration acknowledges support for LAT development, operation and data analysis from NASA and DOE (United States), CEA/Irfu and IN2P3/CNRS (France), ASI and INFN (Italy), MEXT, KEK, and JAXA (Japan), and the K.A.~Wallenberg Foundation, the Swedish Research Council and the National Space Board (Sweden). Science analysis support in the operations phase from INAF (Italy) and CNES (France) is also gratefully acknowledged. This work performed in part under DOE Contract DE-AC02-76SF00515.
\textbf{FACT}: The important contributions from ETH Zurich grants ETH-10.08-2 and ETH-27.12-1 as well as the funding by the Swiss SNF and the German BMBF (Verbundforschung Astro- und Astroteilchenphysik) and HAP (Helmoltz Alliance for Astroparticle Physics) are gratefully acknowledged. Part of this work is supported by Deutsche Forschungsgemeinschaft (DFG) within the Collaborative Research Center SFB 876 "Providing Information by Resource-Constrained Analysis", project C3. We are thankful for the very valuable contributions from E. Lorenz, D. Renker and G. Viertel during the early phase of the project. We thank the Instituto de Astrofísica de Canarias for allowing us to operate the telescope at the Observatorio del Roque de los Muchachos in La Palma, the Max-Planck-Institut für Physik for providing us with the mount of the former HEGRA CT3 telescope, and the MAGIC collaboration for their support.

\bibliographystyle{JHEB}
\bibliography{./mnemonic,./aa_abbrv,./lit_FFFB}

\providecommand{\href}[2]{#2}\begingroup\raggedright\begin{thebibliography}{10}

\bibitem{Abdo_2010_SEDs}
A.~A. {Abdo}, M.~{Ackermann}, I.~{Agudo}, M.~{Ajello}, H.~D. {Aller}, M.~F.
  {Aller} et~al., \emph{{The Spectral Energy Distribution of Fermi Bright
  Blazars}}, \href{https://doi.org/10.1088/0004-637X/716/1/30}{\emph{ApJ}
  {\bfseries 716} (2010) 30} [\href{https://arxiv.org/abs/0912.2040}{{\ttfamily
  0912.2040}}].

\bibitem{Ghisellini_2017}
G.~{Ghisellini}, C.~{Righi}, L.~{Costamante} and F.~{Tavecchio}, \emph{{The
  Fermi blazar sequence}},
  \href{https://doi.org/10.1093/mnras/stx806}{\emph{MNRAS} {\bfseries 469}
  (2017) 255} [\href{https://arxiv.org/abs/1702.02571}{{\ttfamily
  1702.02571}}].

\bibitem{Madejski_Sikora_2016}
G.~G. {Madejski} and M.~{Sikora}, \emph{{Gamma-Ray Observations of Active
  Galactic Nuclei}},
  \href{https://doi.org/10.1146/annurev-astro-081913-040044}{\emph{ARA\&A}
  {\bfseries 54} (2016) 725}.

\bibitem{fermi_4fgl}
S.~{Abdollahi}, F.~{Acero}, M.~{Ackermann}, M.~{Ajello}, W.~B. {Atwood},
  M.~{Axelsson} et~al., \emph{{Fermi Large Area Telescope Fourth Source
  Catalog}}, \href{https://doi.org/10.3847/1538-4365/ab6bcb}{\emph{ApJS}
  {\bfseries 247} (2020) 33}
  [\href{https://arxiv.org/abs/1902.10045}{{\ttfamily 1902.10045}}].

\bibitem{Meyer_2019}
M.~{Meyer}, J.~D. {Scargle} and R.~D. {Blandford}, \emph{{Characterizing the
  Gamma-Ray Variability of the Brightest Flat Spectrum Radio Quasars Observed
  with the Fermi LAT}},
  \href{https://doi.org/10.3847/1538-4357/ab1651}{\emph{ApJ} {\bfseries 877}
  (2019) 39} [\href{https://arxiv.org/abs/1902.02291}{{\ttfamily 1902.02291}}].

\bibitem{2010apsp.conf..681B}
T.~{Bretz} and D.~{Dorner}, \emph{{MARS - CheObs ed. -- A flexible Software
  Framework for future Cherenkov Telescopes}},  in \emph{Astroparticle,
  Particle and Space Physics, Detectors and Medical Physics Applications},
  C.~{Leroy}, P.-G. {Rancoita}, M.~{Barone}, A.~{Gaddi}, L.~{Price} and
  R.~{Ruchti}, eds., pp.~681--687, Apr., 2010,
  \href{https://doi.org/10.1142/9789814307529\_0111}{DOI}.

\bibitem{2019ICRC...36..630B}
M.~{Beck}, A.~{Arbet-Engels}, D.~{Baack}, M.~{Balbo}, A.~{Biland}, M.~{Blank}
  et~al., \emph{{FACT - Probing the Periodicity of Mrk 421 and Mrk 501}},  in
  \emph{36th International Cosmic Ray Conference (ICRC2019)}, vol.~36 of
  \emph{International Cosmic Ray Conference}, p.~630, July, 2019.

\bibitem{2017ICRC...35..779H}
D.~{Hildebrand}, M.~L. {Ahnen}, M.~{Balbo}, A.~{Biland}, T.~{Bretz}, J.~{Buss}
  et~al., \emph{{Using Charged Cosmic Ray Particles to Monitor the Data Quality
  of FACT}},  in \emph{35th International Cosmic Ray Conference (ICRC2017)},
  vol.~301 of \emph{International Cosmic Ray Conference}, p.~779, Jan., 2017.

\bibitem{2021A&A...647A..88A}
A.~{Arbet-Engels}, D.~{Baack}, M.~{Balbo}, A.~{Biland}, M.~{Blank}, T.~{Bretz}
  et~al., \emph{{The relentless variability of Mrk 421 from the TeV to the
  radio}}, \href{https://doi.org/10.1051/0004-6361/201935557}{\emph{A\&A}
  {\bfseries 647} (2021) A88}
  [\href{https://arxiv.org/abs/2101.10651}{{\ttfamily 2101.10651}}].

\bibitem{4LAC_2019}
{The Fermi-LAT collaboration}, \emph{{The Fourth Catalog of Active Galactic
  Nuclei Detected by the Fermi Large Area Telescope}}, {\emph{arXiv e-prints}
  (2019) arXiv:1905.10771} [\href{https://arxiv.org/abs/1905.10771}{{\ttfamily
  1905.10771}}].

\bibitem{Scargle_2013}
J.~D. {Scargle}, J.~P. {Norris}, B.~{Jackson} and J.~{Chiang}, \emph{{Studies
  in Astronomical Time Series Analysis. VI. Bayesian Block Representations}},
  \href{https://doi.org/10.1088/0004-637X/764/2/167}{\emph{ApJ} {\bfseries 764}
  (2013) 167} [\href{https://arxiv.org/abs/1207.5578}{{\ttfamily 1207.5578}}].

\bibitem{Sokolov_2004}
A.~{Sokolov}, A.~P. {Marscher} and I.~M. {McHardy}, \emph{{Synchrotron
  Self-Compton Model for Rapid Nonthermal Flares in Blazars with
  Frequency-dependent Time Lags}},
  \href{https://doi.org/10.1086/423165}{\emph{ApJ} {\bfseries 613} (2004) 725}
  [\href{https://arxiv.org/abs/astro-ph/0406235}{{\ttfamily
  astro-ph/0406235}}].

\bibitem{Giannios_2009}
D.~{Giannios}, D.~A. {Uzdensky} and M.~C. {Begelman}, \emph{{Fast TeV
  variability in blazars: jets in a jet}},
  \href{https://doi.org/10.1111/j.1745-3933.2009.00635.x}{\emph{MNRAS}
  {\bfseries 395} (2009) L29}
  [\href{https://arxiv.org/abs/0901.1877}{{\ttfamily 0901.1877}}].

\bibitem{Burd_2021}
P.~R. {Burd}, L.~{Kohlhepp}, S.~M. {Wagner}, K.~{Mannheim}, S.~{Buson} and
  J.~D. {Scargle}, \emph{{Ornstein-Uhlenbeck parameter extraction from light
  curves of Fermi-LAT observed blazars}},
  \href{https://doi.org/10.1051/0004-6361/202039097}{\emph{A\&A} {\bfseries
  645} (2021) A62} [\href{https://arxiv.org/abs/2010.12318}{{\ttfamily
  2010.12318}}].

\bibitem{Shukla2020}
A.~{Shukla} and K.~{Mannheim}, \emph{{Gamma-ray flares from relativistic
  magnetic reconnection in the jet of the quasar 3C 279}},
  \href{https://doi.org/10.1038/s41467-020-17912-z}{\emph{Nature
  Communications} {\bfseries 11} (2020) 4176}.

\bibitem{ATel_FACT2017}
A.~{Biland}, D.~{Dorner}, J.~A. {Garcia-Gonzalez}, I.~{Martinez}, {FACT
  Collaboration} and {HAWC Collaboration}, \emph{{Mrk 421 showing enhanced
  activity at TeV energies with two short bright outbursts}}, {\emph{The
  Astronomer's Telegram} {\bfseries 11077} (2017) 1}.

\bibitem{Meyer_2021}
M.~{Meyer}, M.~{Petropoulou} and I.~M. {Christie}, \emph{{The Observability of
  Plasmoid-powered {\ensuremath{\gamma}}-Ray Flares with the Fermi Large Area
  Telescope}}, \href{https://doi.org/10.3847/1538-4357/abedab}{\emph{ApJ}
  {\bfseries 912} (2021) 40}
  [\href{https://arxiv.org/abs/2012.09944}{{\ttfamily 2012.09944}}].

\end{thebibliography}\endgroup
\end{document}